\documentclass{elsart}
\usepackage{amssymb}
\usepackage{epsfig}
\begin{document}
\begin{frontmatter}
\title{Skyrme Mean-Field Study of Rotational Bands in Transfermium Isotopes}
\author[ULB]{M. Bender}
\ead{mbender@ulb.ac.be}
\author[Sac]{P. Bonche}
\author[Sac]{T. Duguet\thanksref{newadd}}
\author[ULB]{P.-H. Heenen}

\address[ULB]{Service de Physique Nucl\'eaire Th\'eorique,
         Universit\'e Libre de Bruxelles, C.P. 229, B-1050 Bruxelles,
         Belgium}
\address[Sac]{Service de Physique Th\'eorique,
         CEA Saclay, 91191 Gif sur Yvette Cedex,
         France}
\thanks[newadd]{Current address: 
        Physics Division, Argonne National Laboratory, Argonne, IL 60439}
\date{December 12 2002}
%
%
\begin{abstract}
Self-consistent mean field calculations with the SLy4 interaction and a 
density-dependent pairing force are presented for nuclei in the 
Nobelium mass region. Predicted quasi-particle spectra are compared 
with experiment for the heaviest known odd $N$ and odd $Z$ nuclei. 
Spectra and rotational bands are presented for nuclei 
around $^{252,4}$No for which experiments are either planned or already 
running.
\end{abstract}
\begin{keyword}
\PACS 21.60.Jz \sep 27.90.+b \sep 21.10.Pc \sep 21.10.Re
\end{keyword}
\end{frontmatter}  


\section{Introduction}

Effective interactions used in microscopic mean-field models are built
from nuclear ingredients such as ground-state properties of 
selected nuclei and infinite nuclear matter~\cite{Bhr03}.
No information coming from rotational bands of deformed nuclei 
is a priori implemented in these forces.
The variation of the moment of inertia along a band is related to 
the evolution of pairing correlations with rotation. 
Such a variation brings other constraints on the structure and 
the strength of the pairing interaction~\cite{Dug01}.
Furthermore, properties of bands in odd-mass nuclei are a fingerprint of 
the quasi-particle excitations on which they are built and, thus, of 
the single-particle spectra predicted by the effective mean-field interaction.
Experimental study of rotational spectra in nuclei far from stability 
is thus an important tool to test the predictive power of these forces.

The recent experimental results for the Nobelium isotopes $^{252}$No 
and $^{254}$No~\cite{Rei99,Lei99,Rei00,Her02,But02} 
are in this respect of particular importance. 
They bring data in a region close to the domain of super-heavy nuclei, 
where our knowledge of single-particle spectra and of pairing correlations 
is particularly limited.
Since then, further experiments have been proposed or are already 
underway for other even-even and odd-mass nuclei in this mass region; 
new results are expected to appear soon.
This experimental context has triggered 
a number of theoretical studies with various mean-field models.
Fully self-consistent cranked HFB calculations have been performed 
for selected nuclei with several interactions: the Gogny force~\cite{Egi00}, 
the Skyrme interactions SLy4 and a density-dependent pairing 
force~\cite{Dug01} and SkM* together with a schematic pairing 
force~\cite{Laf01}. The relativistic mean-field model \cite{Afa02} 
has also been applied to this domain. Finally, let us quote the large 
scale microscopic-macroscopic study of Muntian~\etal~\cite{Mun99}.

In this paper, we present calculations for the rotational bands of some 
even-even nuclides in the \mbox{$A=250$} mass region and of odd-mass isotopes 
differing from $^{252}$No or $^{254}$No by either one neutron or one proton. 
We also discuss the spectra of quasi-particle states expected for these nuclei.

First, we briefly recall our method and present its main ingredients. 
We then calculate the spectra and rotational bands of the heaviest 
isotopes for which there are extensive experimental data.
Finally, we calculate spectra and rotational bands of nuclei around 
the No isotopes. 
%
%
\section{The method}
We have performed cranked HFB calculations following 
the method described in ref.~\cite{Ter95}.
The SLy4 parameterization of the Skyrme interaction~\cite{Cha98} is chosen
in the mean-field (particle-hole) channel.
In the pairing (particle-particle) channel, we use a zero-range 
density-dependent force with the strength ($V = - 1250$ MeV~fm$^3$) 
and cutoff (5~MeV) as in our previous study of $^{252,254}$No~\cite{Dug01}.
The Lipkin-Nogami prescription is utilized to
prevent a collapse of pairing correlations at high angular momentum.
Quasi-particle excitations are calculated fully 
self-consistently~\cite{HJ98,Dug02a}. 
The effect of core polarization and of the blocking of a pair of states 
are thus exactly taken into account.

This model has been successfully used to describe super-deformed rotational 
bands~\cite{Rbf99}
and super-heavy nuclei~\cite{Cwi96}. 
In ref.~\cite{Dug01}, we have compared our results 
with data on rotational bands of $^{252}$No and $^{254}$No.
%
%
\section{Results}
\subsection{Single-particle levels}
\begin{figure}[t!]
\centerline{\epsfig{file=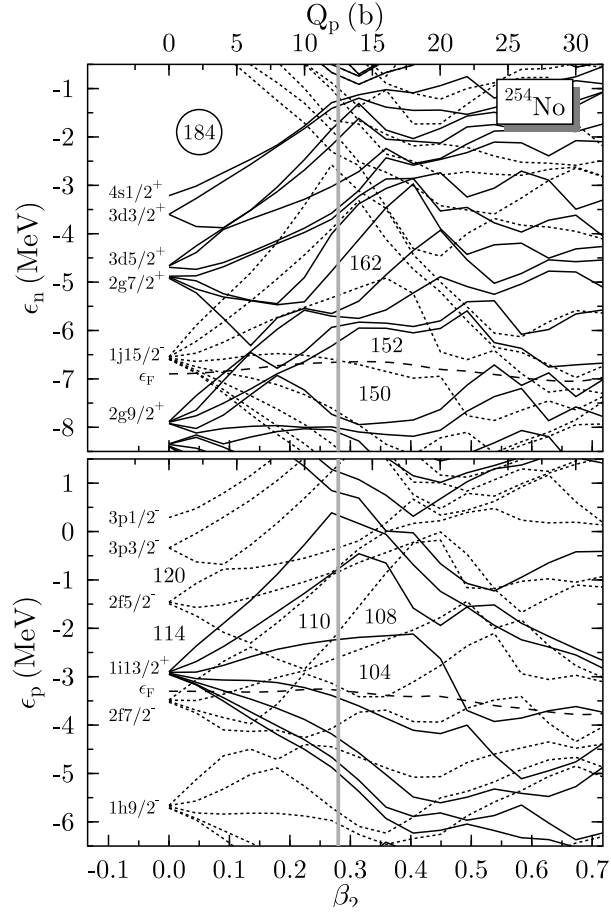}}
\caption{\label{fig:sp:no}
Single-particle energies for neutrons (top) and protons (bottom) calculated 
for $^{254}$No (\mbox{$Z=102$}, \mbox{$N=152$}). 
The Fermi levels are indicated by long-dashed lines and the ground state 
deformation by a vertical gray line.}
\end{figure}

The single-particle levels of $^{254}$No are plotted in figure~\ref{fig:sp:no} 
as a function of the axial quadrupole moment. 
These energies are eigenvalues of the particle-hole part of the 
HFB Hamiltonian. They provide a useful link to simpler models.

The proton level density around the Fermi surface is rather low at 
deformations around \mbox{$\beta_2 = 0.28$} which is the deformation 
that we obtain for $^{254}$No. In this region, there are numerous 
deformed shell closures in the proton spectrum: at \mbox{$Z=98$} and 
100, below $^{254}$No and also at \mbox{$Z=104$}, 108, 110. 
At the same deformation, there are regions of low neutron level density 
around \mbox{$N=150$} and 152 as well as \mbox{$N=162$}.
%
%
\begin{figure}[t!]
\centerline{\epsfig{file=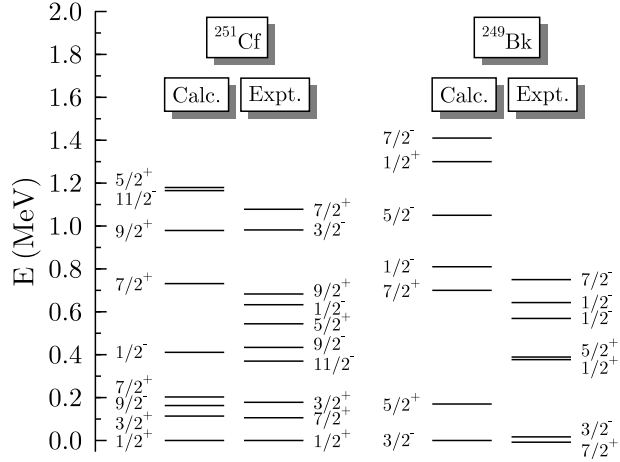}}
\caption{\label{fig:spectra:expt}
Low lying energy spectra of $^{251}_{\ 98}$Cf$_{153}$ 
(odd $N$) and $^{249}_{\ 97}$Bk$_{152}$ (odd $Z$).}
\end{figure}
\subsection{Quasi-particle states in $^{251}$Cf and $^{249}$Bk}
Before applying our method to nuclei for which experimental data are sparse, 
let us focus on the spectra of $^{251}$Cf and $^{249}$Bk 
which are the heaviest odd-neutron and odd-proton nuclei 
for which the low energy spectra are rather well known.
Figure~\ref{fig:spectra:expt} gives the spectra of these odd nuclei 
calculated self-consistently from 
one quasi-particle excitations (1qp) built on an HFB vacuum.
All states which are calculated to lie below 1.2~MeV are 
compared with experiment.
Data are from ref.~\cite{Art99} and references therein for both nuclei and 
also from ref.~\cite{Ahm00} for $^{251}$Cf.

The states we have calculated in $^{251}$Cf are characterized by 
a deformation $\beta_2$ 
around 0.26, with the exception of the $1/2^-$ state at 410~keV, which 
is significantly more deformed ($\beta_2=0.273$ using the relation 
between $\beta_2$ and the quadrupole and hexadecapole moments 
given in ref.~\cite{Cwi96}). 
These values are significantly larger than the fixed $\beta_2$ value used by 
{\'C}wiok~\etal~\cite{Cwi94} (0.247) in their calculations based on 
a Woods-Saxon potential and the Strutinsky prescription. 
However, a clear conclusion is difficult to draw: in ref.~\cite{Cwi94}, 
$\beta_2$ is \emph{a priori} fixed to be the same for all levels, whereas 
in self-consistent mean-field calculations, deformation is optimized 
level per level. In addition, the 1qp energies given in~\cite{Cwi94} 
are eigenvalues of the HFB Hamiltonian using an average potential, 
while our 1qp energies are total binding energy differences between 
self-consistent one quasi-particle states.

The ground state of $^{251}$Cf is obtained from the 1qp excitation 
predominantly based on the first single-particle state above the Fermi level,
which has a spin of $1/2^+$, in agreement with the data. 
The first three excited levels correspond to 1qp states based on the 
single-particle levels just above ($3/2^+$ and $7/2^+$) or just below 
($9/2^-$) the Fermi level. This result is also consistent with experiment, 
with the largest error on the $9/2^-$ state, too low by almost 300~keV. 
At higher excitation energies, the energy of
the $11/2^-$ state is largely overestimated by our calculation. 
Other states are predicted with an accuracy of the order of 300~keV. 
We also obtain a $7/2^+$ state at 700~keV. 
Experimentally, the $3/2^-$ state at 981~keV and the $7/2^+$ state at 1078~keV 
have been interpreted as due to the coupling of the $7/2^+$ 1qp excitation 
with the $2^-$ octupole vibration and the $0^+$ pair vibration respectively.
These states are thus outside the scope of our calculation.
No $3/2^-$ single-particle state is obtained below 1.2~MeV in our calculation. 
A rather similar spectrum is obtained in ref.~\cite{Cwi94}, with however 
a much better energy for the $11/2^-$ state at~370 keV. The RMF results 
of ref.~\cite{Afa02} are in similar agreement with the data than ours, 
with, however a less dense spectrum for the two interactions that were tested. 

We obtain a spin $3/2^-$ for the ground state of $^{249}$Bk.
The largest discrepancy with experiment isthe much too high energy for 
the $7/2^+$ state which is experimentally nearly 
degenerate with the ground state. Our calculation also
significantly overestimates the energy of the first $1/2^+$ state.
All states have a very similar deformation 
$\beta_2$ around 0.255. This value is slightly larger than the fixed 
0.243 value used in~\cite{Cwi94}. The spin obtained 
in~\cite{Cwi94} for the ground state is $3/2^-$, but the $7/2^+$ 
state is excited by only 150~keV. The RMF calculation~\cite{Afa02} 
is in slightly better agreement with the data than our results, mainly 
because of a better energy for the $7/2^+$ state.

It is tempting to interpret the discrepancies observed between the
calculated and experimental spectra in terms of the ordering of the
single-particle levels at sphericity.
For the neutrons and the $^{251}$Cf spectrum, the too low 1qp energy 
of the $9/2^-$ state above the Fermi level and the too high energy of 
the $11/2^-$ state below the Fermi level could be corrected by a lowering 
of the energy of the $1j15/2^-$ sub-shell.
This would increase the \mbox{$N=164$} spherical energy gap, and also 
slightly reduce the deformed gap at \mbox{$N=150$} in favor of an increased 
\mbox{$N=152$} gap for typical ground state deformations around 
\mbox{$\beta = 0.28$}. 
The fact that the self-consistent models underestimate the \mbox{$N=152$} 
shell effect is also suggested by the
experimental two-neutron separation energies~\cite{Bue98a} 
and $Q_\alpha$ values~\cite{Cwi99,Ben00}. 

The positions of the $5/2^+$ and $7/2^+$ states in $^{249}$Bk 
would be in better agreement with the data
if the $1i13/2^+$ shell were lower and
if the splitting between the $3/2^-$ and the $1/2^-$ 
orbitals (labeled as $[521]3/2^-$ and $[521]1/2-$ Nilsson states) 
was reduced by closer spherical $2f7/2^-$ and $2f/5/2^-$ shells.
A lowering of the energies of the $1j15/2^-$ neutron and $1i13/2^+$ 
proton sub-shells can be achieved by an increase of the spin-orbit strength.
However, this is not consistent with analyses showing that most 
Skyrme interactions overestimate the spin-orbit splitting in heavy 
nuclei~\cite{Ben99a}.
In any case, to improve the description of the spectra by modifying the
effective force requires deeper 
modifications than just the spin-orbit strength. One must also note
that slight shifts of the levels could  result from 
deformation modes not included in our model, like the octupole mode.

Such modifications of the spherical sub-shells would affect the shell gaps
for super-heavy nuclei. It would lead to a reduction of the deformed 
\mbox{$Z=110$} gap and an increase of the \mbox{$Z=108$} gap, in better
agreement with the sparse available data.
At the same time, the modified $1i13/2^+$ and $2f7/2^-$ sub-shells would 
affect the \mbox{$Z=114$} gap in opposite ways and might not change it 
quantitatively. Note that most parameterizations of the RMF model predict 
the proton $1i13/2^+$ sub-shell to be below the $2f7/2^-$ 
state~\cite{Ben99a,Afa02}. However, speculations about the magicity of 
\mbox{$Z=114$} and \mbox{$Z=120$} based solely on the deformed one 
quasi-particle states in $^{249}$Bk is questionable
as increasing the charge number polarizes the nuclear density 
distribution, which feeds back into the single-particle spectra. 
For example, the spin-orbit splitting of $2f$ and $3p$ levels nearly
vanishes for nuclides with semi-bubble shapes~\cite{Dec99a} that are
predicted to appear around $^{292}$120~\cite{Ben99a}.

One must also keep in mind that  large changes in the single-particle 
spectra will also affect other regions of the mass table.
In particular, the lower members of the intruder proton $1i13/2^+$ and
neutron $1j15/2^-$ orbitals are responsible of several super-deformed
rotational bands in the \mbox{$A=190$} mass region and cannot be 
changed by more than one or two hundred keV~\cite{HJ98}. 

Recently, conversion electron cascades have been measured 
by Butler \etal~\cite{But02}. The presence of  high K-isomers was
conjectured to explain the observed  intensities.
We have calculated the  2qp-excitations in $^{254}$No which can
lead to high K-states. Most of them have excitation energies significantly
larger than 2.0~MeV. The 2qp $7^-$ state based on $7/2^+$ and $7/2^-$ 
particle states (see figure~\ref{fig:sp:no}) is the only excitation that
our calculation gives below 2.0~MeV, around 1.5~MeV. 
We have also determined its magnetic moment following the method 
developed by Hamamoto~\cite{Ham78}. This configuration has a very large
magnetic moment, around 1.5~W.U., as do several of the more excited
configurations. Whether such a result is consistent with the present 
analysis of the experimental data requires further investigation. 
%
%
\subsection{Pu isotopes}
\begin{figure}[t!]
\centerline{\epsfig{file=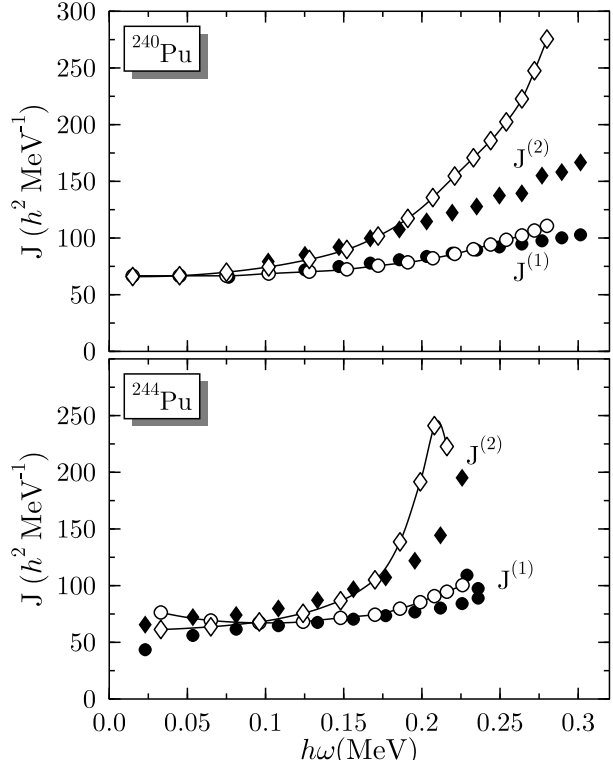}}
\caption{\label{fig:j:pu}
Kinematical (circles) and dynamical (diamonds) moment of inertia for 
$^{240}$Pu (top) and $^{244}$Pu (bottom). 
Open (filled) markers denote calculated (experimental) values.}
\end{figure}
Here, we discuss results of our method for the rotational bands
of $^{240}$Pu and $^{244}$Pu for which recent detailed
experimental data~\cite{Hac98,Wie99,Jan02} have become available. 
The kinematic ($J^{(1)}$) and dynamic ($J^{(2)}$) moments of inertia
are compared with experiment in figure~\ref{fig:j:pu}.

In both cases, the $J^{(1)}$ moment is in  close agreement with the data 
at all rotational frequencies, while $J^{(2)}$ is overestimated at high 
rotational frequencies  in a more pronounced way 
for $^{240}$Pu than for $^{244}$Pu.
In $^{244}$Pu, the calculation slightly underestimates the rotational 
frequency at which the alignment of a proton $i13/2^+$  occurs. 
The same alignment does not occur in $^{240}$Pu up to \mbox{$J=34$}~$\hbar$,
both experimentally and theoretically.
It has been suggested that this absence is due to
octupole correlations~\cite{Wie99} which however are not included 
in our calculation. Further investigations are necessary to clarify 
this point.
%
%
\subsection{Quasi-particle states of nuclei around $^{252,254}$No}
\begin{figure}[t!]
\centerline{\epsfig{file=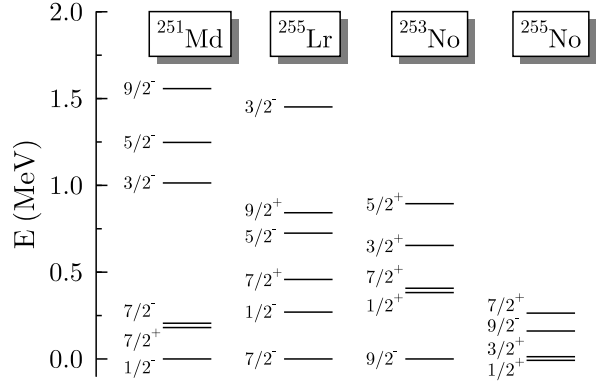}}
\caption{\label{fig:spectra:odd}
Low lying energy spectra of $^{251}_{101}$Md,
$^{255}_{103}$Lr (odd $Z$), $^{253}_{102}$No$_{151}$ and 
$^{255}_{102}$No$_{153}$ (odd $N$).}
\end{figure}
The spectra of four selected odd-mass nuclei 
around $^{252,254}$No are presented on figure~\ref{fig:spectra:odd}.
All states are obtained by self-consistent 1qp excitation on a HFB vacuum.
The experimental data on these nuclei are sparse~\cite{Art99}.

Let us focus first on the two odd-proton nuclei for which there are no data.
As expected from the single-particle spectrum of figure~\ref{fig:spectra:expt},
the ground state of $^{251}$Md has a calculated spin $1/2^-$, with two 
closely packed $7/2$ states of opposite parity around 200~keV. 
For $^{255}$Lr, the three same spins appear at low energy,
with a $7/2^-$ ground state and two excited states below 0.5~MeV.
Three more states than in $^{251}$Md are obtained below 1.0~MeV.
This is a consequence of the denser single-particle spectrum above $Z=104$.
{\'C}wiok~\etal~\cite{Cwi94} have obtained similar but less dense
spectra, with the same spins for the ground states. 
The main difference relies  in the position of $9/2^+$ states. 
In $^{255}$Lr, we obtain a state at an energy around 600~keV higher
than in ref.~\cite{Cwi94}. In $^{251}$Md, we do not obtain it 
below 1.5~MeV. These differences are probably related to the position 
of the $i13/2^+$ sub-shell discussed above.

The ground state of $^{253}$No is predicted to be a $9/2^-$ level, in 
agreement with the tentative experimental assignment~\cite{Art99}. 
Note that this state arises from the $i13/2^+$ sub-shell. 
A lowering of this sub-shell by more than a
few 100~keV will affect the ordering of the $^{253}$No first levels.
The spectrum that we obtain is in overall 
good agreement with the results of ref.~\cite{Cwi94}. 
In particular and contrary to the spectrum reported in ref.~\cite{Art99}, 
we do not obtain a $11/2^-$ state at low energy. However, 
our calculation predicts a $1/2^+$ state at 300~keV.
We have also calculated the magnetic moments of these states
which turn out to be rather large (1~W.U. for the $9/2^-$, 
0.25~W.U. for the $1/2^+$ and 0.1~W.U. for the $7/2^+$).
Hence, low lying states of rotational bands build on these excitations
will deexcite predominantly via $M1$ transitions. The ground state of 
$^{255}$No is a $1/2^+$ state, nearly degenerate with a $3/2^+$ state. 
Two other states are predicted to lie below 300 keV, as can be expected 
from the neutron single-particle spectrum of figure~\ref{fig:sp:no}.
%
%
\subsection{Rotational bands of nuclei around $^{252,254}$No}
\begin{figure}[t!]
\centerline{\epsfig{file=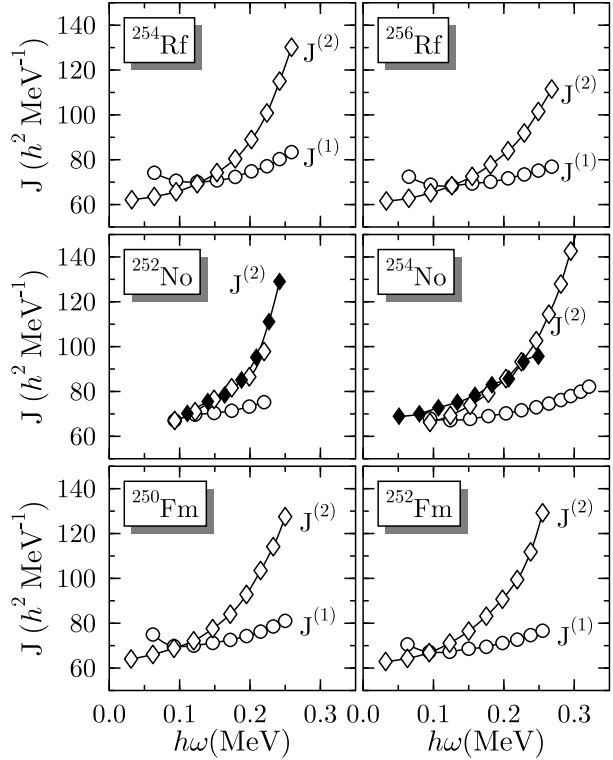}}
\caption{\label{fig:j:even}
Calculated kinematic, $J^{(1)}$, and dynamic, $J^{(2)}$, 
moments of inertia for nuclides with  \mbox{$Z=100$}--104
and \mbox{$N=150$}, 152.
Empty symbols are for calculations, full ones for experiment.}
\end{figure}
Calculated moments of inertia for the even-even nuclei with 
\mbox{$Z=100$}--104 and \mbox{$N=150$}, 152 are shown in 
figure~\ref{fig:j:even}. Experimental data for $^{252}$No are taken 
from~\cite{Rei99}, data for $^{254}$No from~\cite{Her02}. 
The overall behavior of the moments of inertia of 
the Fm and Rf isotopes is very similar to the one obtained 
previously in ref.~\cite{Dug01} for $^{252}$No and $^{254}$No. 
The predicted alignment for $^{252}$No underestimates the data,
while for $^{254}$No  calculations and experiement are in reasonable 
agreement. The heavy \mbox{$N=152$} isotones $^{254}$No and $^{256}$Rf
have a significantly slower alignment than the other nuclides, with
a minimum predicted for $^{254}$No. 
Following our previous discussion on the location of the high $j$ states, 
we observe that an energy decrease for the $j15/2^-$ orbital would result 
in a faster alignment in $^{252}$No and a slower one in $^{254}$No,
as can be inferred from the quasi-particle energies given 
on figure~4 in ref.~\cite{Dug01}.
\begin{figure}[t!]
\centerline{\epsfig{file=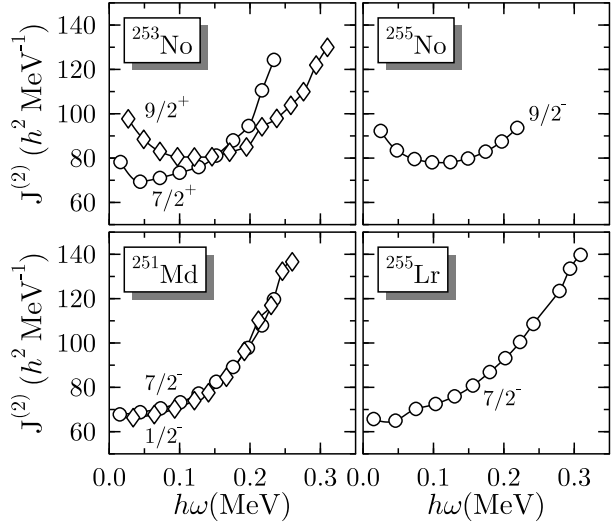}}
\caption{\label{fig:j:odd}
Calculated dynamic moments of inertia $J^{(2)}$ for
$^{253}_{102}$No$_{151}$, $^{255}_{102}$No$_{153}$,
$^{251}_{101}$Md$_{150}$ and $^{255}_{103}$Lr$_{152}$ 
versus rotational frequency.}
\end{figure}

The moment of inertia of selected bands in odd-mass nuclei are plotted in 
figure~\ref{fig:j:odd}.
The odd proton isotopes have rather similar moment of inertia, the $1/2^-$
and $7/2^-$ bands in $^{251}$Md behaving in essentially the same way.
%
%
%
\section{Conclusions}
In this work, we have compared our self-consistent mean-field 
calculations with the heaviest two odd nuclei for which the 
single-particle spectrum is reasonably known, $^{251}$Cf and $^{249}$Bk.
The qualitative overall agreement is rather good, in spite of some 
discrepancies for a few states.
This gives us confidence in our results for the other, 
less known, nuclei that we studied in this mass region. 
In particular, our results may be of some value for 
experiments intended to explore any of these nuclei.
However, one should bear in mind that these states have been 
calculated with the HFB formalism. 
There is, therefore, no guarantee that they are orthogonal, 
especially when corresponding to the same parity and $K$ value.

Mean-field models~\cite{Cwi96,Ben99a,Rut97a,Kru00a} predict different 
neutron and proton numbers for spherical, doubly magic super-heavy nuclei.
The mass region we have explored is the heaviest one for which one has 
sufficient experimental knowledge to extract some information  relevant
to the single particle spectra.
However, this information and its consequence on magic numbers 
cannot be extrapolated without care from $^{254}$No to higher masses.
For instance, deformation of the density arising from multipole moments 
higher than the quadrupole one are automatically taken into account 
in microscopic mean-field calculations. 
As these higher moments change rapidly with nucleon numbers, 
substantial modifications in the single-particle spectrum 
and thus of magic numbers may be expected (see e.g.\ ref.~\cite{Pat91}).

Another interesting finding is the sensitivity of the calculated 
spectra of odd nuclides to the precise location of individual orbitals.
In these very heavy nuclei, the density of levels is higher than 
for lower masses.
One may thus speculate that a slight modification of the spin-orbit strength 
could result in some significant changes in excitation spectra 
in this mass region while remaining of no importance for lighter nuclei.
This points to the necessity of a careful analysis of the spin orbit strength 
in standard fit procedure of effective forces, 
which is beyond the scope of the present work.
%
%
\section*{Acknowledgements}
This research was supported in part by the PAI-P5-07 of the Belgian
Office for Scientific Policy. M.~B.\ acknowledges support through a 
European Community Marie Curie Fellowship.
We thank R.-D.\ Herzberg, R.\ V.\ F.\ Janssens, T.~L.\ Khoo, 
W.\ Nazarewicz, Ch.\ Theisen for fruitful and inspiring discussions. 
%
%

\end{document}